\documentclass[preprint,journal]{vgtc}            

\usepackage{mathptmx}                  
\usepackage{textcomp}

\usepackage{amsmath}
\usepackage{subfig}

\onlineid{1597}

\vgtccategory{Research}

\vgtcpapertype{Technical}

\title{Perceptual Thresholds for Radial Optic Flow Distortion\\in Near-Eye Stereoscopic Displays}

\author{%
\authororcid{Mohammad R. Saeedpour-Parizi}{0000-0001-7820-2495}, \authororcid{Niall L. Williams}{0000-0002-0273-883X}, \textit{Student Member, IEEE},\\\authororcid{Tim Wong}{0009-0004-2589-0377}, \authororcid{Phillip Guan}{0000-0001-6966-4110}, \authororcid{Dinesh Manocha}{0000-0001-7047-9801}, \textit{Fellow, IEEE}, and \authororcid{Ian M. Erkelens}{0000-0002-8384-2777}
}

\authorfooter{
  \item
  	Mohammad R. Saeedpour-Parizi, Tim Wong, Phillip Guan, and Ian M. Erkelens are with Meta Reality Lab Research.
  	E-mail: \{rezasaeedpour$\vert$tlwong$\vert$philguan$\vert$ian.erkelens\}@meta.com
  \item
  	Niall L. Williams and Dinesh Manocha are with the University of Maryland, College Park.
  	E-mail: \{niallw$\vert$dmanocha\}@umd.edu
}

\abstract{
We provide the first perceptual quantification of user's sensitivity to radial optic flow artifacts and demonstrate a promising approach for masking this optic flow artifact via blink suppression.
Near-eye HMDs allow users to feel immersed in virtual environments by providing visual cues, like motion parallax and stereoscopy, that mimic how we view the physical world.
However, these systems exhibit a variety of perceptual artifacts that can limit their usability and the user's sense of presence in VR.
One well-known artifact is the vergence-accommodation conflict (VAC).
Varifocal displays can mitigate VAC, but bring with them other artifacts such as a change in virtual image size (radial optic flow) when the focal plane changes.
We conducted a set of psychophysical studies to measure users' ability to perceive this radial flow artifact before, during, and after self-initiated blinks.
Our results showed that visual sensitivity was reduced by a factor of 10 at the start and for $\sim \!\! 70$ ms after a blink was detected. 
Pre- and post-blink sensitivity was, on average, $\sim \! 0.15\%$ image size change during normal viewing and increased to $\sim \!\! 1.5-2.0\%$ during blinks. 
Our results imply that a rapid (under 70 ms) radial optic flow distortion can go unnoticed during a blink. 
Furthermore, our results provide empirical data that can be used to inform engineering requirements for both hardware design and software-based graphical correction algorithms for future varifocal near-eye displays.
Our project website is available at \href{https://gamma.umd.edu/ROF/}{\textcolor{blue}{\texttt{https://gamma.umd.edu/ROF/}}}.
} 


\teaser{
  \centering
  \includegraphics[width=\linewidth, alt={A visualization of our experimental set up used to measure users' visual sensitivity to radial optic flow. Participants viewed two sequentially-presented intervals where we rendered a virtual scene on the 3D display with or without radial optic flow artifacts, and they were tasked with identifying the interval containing the motion artifact.}]{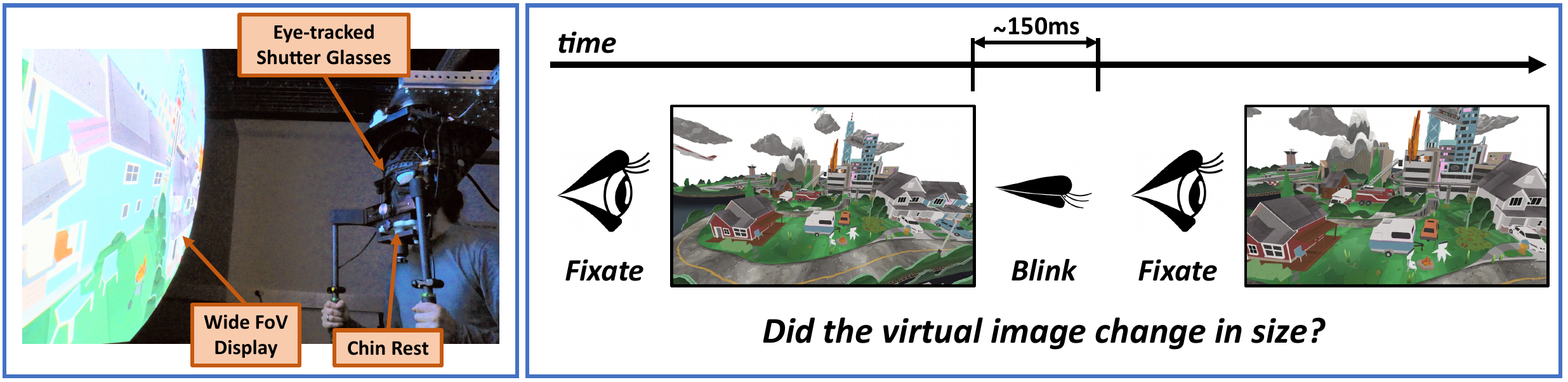}
  \caption{A diagram of our experimental setup used to measure participants' sensitivity to radial optic flow artifacts, which are an unwanted side effect of varifocal head-mounted displays. 
  \textbf{Left:} The participant sits on a height-adjustable stool and places their chin on a chin rest to keep their head steady while viewing the VR display simulator. The participant views the wide field-of-view display through shutter glasses so that they can view the virtual content in 3D. 
  \textbf{Right:} A visualization of \textcolor{black}{one interval in our 2-interval forced choice experiment that participants completed.} \textcolor{black}{In the experiment, participants viewed two sequentially-presented intervals where we rendered a virtual scene on the 3D display. One interval showed the content with the radial optic flow artifact while the other had no change in the rendered image, and participants were tasked with detecting which interval contained the optic flow artifact.} The \textcolor{black}{optic flow stimulus was} implemented as a uniform scaling of the rendered image's size, which the participant perceives as a radial optic flow field. In order to study the effects of blink suppression on participants' sensitivity to radial optic flow, the image changed size before, during, or after voluntary eye blinks (randomly varied from trial to trial).}
  \label{fig:teaser}
}

\graphicspath{{figs/}{figures/}{pictures/}{images/}{./}} 

\usepackage{tabu}                      
\usepackage{booktabs}                  
\usepackage{lipsum}                    
\usepackage{mwe}                       

\begin{document}



\maketitle

\section{Introduction}
\label{sec:introduction}

Virtual reality (VR) displays allow users to interact with computer-generated environments in a way that is natural and immersive.
Unique to VR is the feeling of presence, wherein the user feels that they are truly in the virtual environment that they are perceiving through the head-mounted display (HMD) \cite{slater1997framework}.
Some important factors that contribute to this feeling of presence are a high-resolution, high refresh rate stereoscopic display \cite{hendrix1996presence,barfield1995effect}; a wide field of view (FOV) \cite{lin2002effects}; and low-latency position tracking to render perspective-correct views of the virtual content \cite{hendrix1996presence}.
These features improve the level of immersion provided by the HMD \cite{slater1997framework}, and poor implementations of these features may introduce visual artifacts such as the screen door effect \cite{lee2020key}, screen tearing and flickering, chromatic aberration \cite{zhan2020practical}, and motion artifacts.

Although these artifacts can break a user's sense of presence in the virtual experience, limits of the human visual system (HVS) mean that beyond a certain threshold, these artifacts are imperceptible \cite{gescheider2013psychophysics} (though they may still impact the user experience \cite{goettker2020differences}).
For example, prior work suggests that a total system latency of $50-70$~ms is tolerable for gaze-contingent foveated rendering \cite{albert2017latency}, a measurement that informs HMD manufacturers on how responsive the eye tracking and rendering systems need to be.
Thus, in order to understand how good the display, optical, and rendering systems of an HMD need to be to produce a sufficiently-immersive experience, it is important that we study the perceptual thresholds of the HVS for different types of visual artifacts.

In this work, we mainly study human sensitivity to radial optic flow patterns in wide FOV stereoscopic displays, with applications to distortion correction for varifocal HMDs.
Varifocal HMDs are of interest because they are able to provide accurate focus cues as the user shifts their gaze around the scene to look at objects at different depths.
This ability to provide accurate focus cues mitigates the vergence-accommodation conflict (VAC), a problem which can reduce image quality and cause eye strain and fatigue after prolonged use of the display \cite{kramida2015resolving,hoffman2008vergence}.
One side effect of varifocal HMDs is image distortion that occurs when the system's focal power changes.
In particular, changes in the focal power cause the virtual image size to change (i.e., image \textit{magnification} or \textit{minification}).
To the user, this is perceived as a motion artifact in the form of \textit{radial optic flow} (i.e., the retinal image appears to expand/contract and objects appear to move closer to/further away from the observer), and can manifest as a sensation of self-motion in the world.
To eliminate this artifact, we can apply the inverse of this distortion to the rendered image so that the two distortions (from the applied distortion and the change in focal power) cancel each other out and there is no perceived distortion \cite{lavalle2023virtual,bax2002real}.
In order to know how accurate the inverse distortion needs to be to prevent perception of the radial flow artifact, we need to understand how sensitive users are to radial optic flow.

\textbf{Main Contributions:} We conducted an experiment to measure human observers' sensitivity to expanding radial optic flow in a wide FOV stereoscopic display, motivated by the advent of varifocal HMDs with focus-tunable optics.
To understand the viability of masking the flow artifact during blinks, we measured this sensitivity before, during, and after self-initiated blinks.
Our experiment makes use of recent developments in HMD distortion calibration \cite{guan2022perceptual} and adaptive psychophysics \cite{owen2021adaptive}.
In particular, we used a stereoscopic, wide field-of-view, eye-tracked display simulator \cite{guan2022perceptual} to emulate the radial optic flow distortion effect in a well-controlled environment.
Since our experiment modified both the magnitude of the optic flow effect \textit{and} the delay in the flow onset (relative to blink timing), we used an adaptive psychophysical paradigm \cite{owen2021adaptive} to reduce the number of trials needed in our experiment.
In a psychophysical user study, we collected eye tracking data and subjective perceptual responses to radial optic flow from ten participants.
Our results show that users are extremely sensitive to image magnification artifacts during normal viewing of the virtual content, and that blinks can be an effective method of masking this artifact that yields a $10\times$ decrease in sensitivity to radial optic flow compared to without blink masking.
In particular, we find that:
\begin{itemize}
    \item Visual sensitivity to image magnification during normal viewing (no blink suppression) is about $0.1\% - 0.2\%$ change in image size.
    \item The maximum image magnification that can be hidden during a blink without the user noticing is about $2\%$, with some inter-observer variability ($1.14\% - 2.54\%$).
    \item Visual sensitivity to radial optic flow begins to recover $\sim \!70$~ms after the first detected frame of the start of a blink, and $\sim \!50$~ms before the first detected frame of the end of a blink.
    \item Our results have implications for varifocal HMD design and engineering. Results suggest that the error margins on radial optic flow correction in normal viewing conditions are \textit{very} low and that discrete varifocal systems are an attractive paradigm for implementing varifocal HMDs, since the radial flow artifact can be reliably masked using blinks, as long as it occurs over a short enough time window.
\end{itemize}

\section{Background \& Related Work}
\label{sec:background}

\subsection{Vergence-Accommodation Conflict \& Varifocal Displays}
\label{subsec:VAC}
In order to deliver an immersive virtual experience, HMDs try to recreate the perceptual experience of viewing a 3D scene in the real world, which relies in part on being able to perceive depth \cite{lueder20123d}.
Most HMDs are able to support depth cues like motion parallax, disparity, perspective foreshortening, occlusion, texture gradients, and vergence.
One cue that is currently not supported in existing commercial HMDs is accommodation, wherein the eye changes optical power to maintain a clear image of the gaze target \cite{toates1972accommodation,bruce2003visual}.
Inability to support this depth cue produces the vergence-accommodation conflict (VAC).

\begin{figure*}[t]
    \centering
    \includegraphics[width=.85\textwidth]{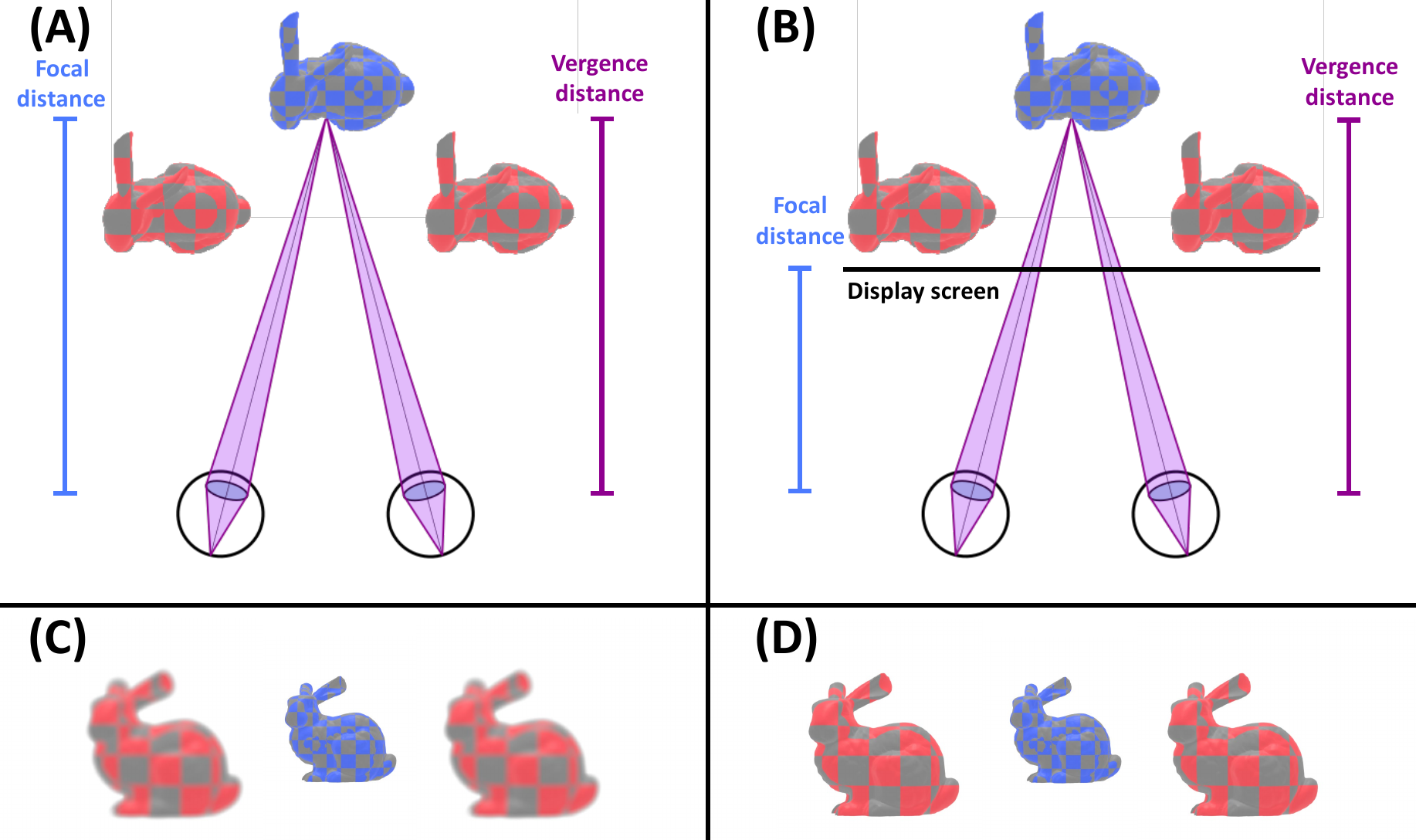}
    \caption{A visualization of the vergence-accommodation conflict (recreated from \cite{hoffman2008vergence}), which varifocal HMDs are designed to mitigate. \textbf{(A)} In real-world viewing conditions, when the observer focuses on an object, their vergence distance and focal distance match. \textbf{(B)} In VR HMD viewing conditions, when the observer focuses on a virtual object rendered on the stereoscopic display, their vergence distance matches that of the real-world condition, but their focal distance is tuned to the distance of the display, which \textit{does not} match the perceived distance of the object. \textbf{(C)} When the vergence and focal distances match, the object being looked at appears in focus and is clear (blue bunny), while objects located at other distances from the observer appear blurred (red bunnies on either side). \textbf{(D)} When the vergence and focal distances do not match (and the observer focuses on a far-away object), all virtual objects appear in focus and are clear due to the fixed focal power of the display. If the user instead looked at an object at a nearby object, all of the virtual content would appear out of focus.}
    \label{fig:vac}
\end{figure*}

VAC is caused by the HMD design: the display is at a fixed distance from the user's eyes and the lenses of the display are calibrated to a fixed focal distance.
Disparity cues from stereoscopic rendering allow the user to rotate (verge) their eyes to look at virtual objects at different distances, but the fixed focal distance prevents the their eyes from accommodating to the perceived depth of the virtual content~\cite{kramida2015resolving}.
This produces conflicting information between vergence and accommodation cues (see \autoref{fig:vac}).
As a result, users experience decreased binocular fusion accuracy \cite{bharadwaj2009accommodative,hoffman2008vergence} and may also experience visual discomfort and fatigue \cite{hoffman2008vergence}.
\textcolor{black}{Furthermore, results from Erkelens et al. suggest that VAC may also degrade users' perception of the real world in augmented reality viewing conditions \cite{erkelens202019}.}

To mitigate this problem, researchers have developed HMDs that support accommodation \cite{kramida2015resolving}.
In this work, we focus on varifocal near eye displays, which support accommodation by changing the focal plane of the display according to the vergence distance of the user.
The focal plane can be changed using one of a number of different techniques, such as mechanically sliding optics, deformable membrane mirrors, tunable lenses, or birefringent lenses.
Regardless of the mechanism used, all varifocal displays suffer from an image magnification (or minification) artifact wherein the virtual image is magnified (or minified) as the focal distance changes \cite{piszczek2023compensation}. 
To the user, this artifact is perceived as the virtual content getting larger or smaller (depending on the direction of the focal power change), which produces a \textit{radial optic flow} field on the retina (see \autoref{fig:teaser}).
To remove this artifact, we can apply the inverse distortion to the rendered image such that the two distortions cancel out and the user perceives an undistorted image.

\subsection{Psychophysics \& Sensitivity to Radial Optic Flow}
\subsubsection{Psychophysics}
Psychophysics is the study of the relationship between physical stimuli and the perceptual response they trigger in observers \cite{gescheider2013psychophysics}.
Psychophysics researchers are usually interested in measuring perceptual thresholds, which describe the level of stimulus intensity required for an observer to perceive the stimulus.
Another common measurement is just-noticeable differences, which describe how much a stimulus must change by in order for the observer to notice that it has changed in intensity \cite{luce1958derivation}.

During a psychophysical experiment, the participant is exposed to the stimulus and then answers a question that provides information on whether or not they perceived the stimulus.
In order to collect reliable results, participants usually complete this task hundreds of times at varying stimulus intensities, and a \textit{psychometric function} that models their perceptual response is fit to the trial response data \cite{wichmann2001psychometric}.
The order and level of the stimulus intensity across trials varies depending on the experiment method used.
In the method of constant stimuli (MCS), the stimulus is set at pre-determined levels, usually in regular intervals, and each level is tested multiple times in a random order \cite{wichmann2001psychometric}.
The method of adjustment (MoA) requires the participant to adjust the stimulus level in real time until it is just noticeable or just becomes unnoticeable \cite{ehrenstein1999psychophysical}.
The method of limits starts at a stimulus level greatly above or below the threshold, and monotonically increases or decreases across trials until the participant detects the stimulus \cite{ehrenstein1999psychophysical}.
The threshold is then calculated as the average of the stimulus level across all termination points.
Adaptive methods, such as the staircase method, adjust the stimulus level according to the user's responses, with the aim of quickly finding the threshold level and staying at that level for a number of trials \cite{ehrenstein1999psychophysical}.
The parameter estimation by sequential testing (PEST) method uses maximum-likelihood estimation to set the stimulus to the most efficient level for a given trial \cite{lieberman1982microcomputer}.
The staircase and PEST methods are similar, except that the change in stimulus level between trials is constant in the staircase method, but varies from trial to trial with the PEST method.
For experiments where the sensitivity to multiple different stimulus parameters (e.g., stimulus intensity, duration, and direction), Owen et al. \cite{owen2021adaptive} developed a nonparametric method for modeling the psychometric function and an active sampling policy to more efficiently sample stimulus parameter values that are close to the observer's threshold.
Similarly, Keely et al.~\cite{keeley2023semi} model the psychometric function using a semi-parametric function with Gaussian priors to further improve the sampling efficiency of the threshold estimation process.
\textcolor{black}{Our work uses the adaptive sampling scheme by Owen et al. \cite{owen2021adaptive} to determine stimulus parameter values (\autoref{subsubsec:exp_stimulus}) from trial to trial, so that we can converge towards each participant's perceptual threshold with fewer trials than would be required of sampling paradigms in traditional psychophysical approaches.}

\subsubsection{Sensitivity to Radial Optic Flow}
Optic flow refers to the pattern of motion perceived by an observer when there is relative motion between the observer and their surrounding environment \cite{gibson1950perception}.
Humans use optic flow as a signal for self-motion by parsing the perceived flow field into rotational and translational components \cite{lappe1999perception}.
That is, different optic flow patterns indicate different types of movement through the environment and sensitivity to these different patterns varies.
Humans rely on optic flow so much for self-motion perception that it is possible for a stationary observer to erroneously feel like they are moving through an environment if they are exposed to the appropriate optic flow stimuli (a sensation known as \textit{vection}) \cite{kooijman2023measuring}.
In the context of near-eye displays, it is important that we build devices that do not create motion artifacts like incorrect optic flow since they may increase the likelihood that the user will experience vection, which can lead to significant feelings of motion sickness \cite{kennedy2010research,bonato2008vection,hettinger1992visually}.

In order to know how much correction needs to be applied in the rendering pipeline to make the image magnification artifact of varifocal displays imperceptible, it is important that we first measure how well users can perceive radial optic flow patterns. 
Researchers have studied perceptual sensitivity to optic flow patterns under a variety of conditions and have found evidence that the brain has specialized detectors for different patterns of optic flow (radial, translational, and rotational) \cite{freeman1992human,regan1985visual,morrone1995two,snowden1996effects,snowden1997phantom}.
Freeman et al. \cite{freeman1992human} measured how sensitive observers are to sinusoidally expanding and contracting radial flow fields and found a threshold of about $0.075^\circ$.
Joshi et al. \cite{joshi2015development} measured the motion coherence threshold for radial optic flow and found that observers required a higher proportion of motion coherence for lower flow speeds ($1.6^{\circ}/\text{s}$) compared to higher flow speeds ($5.5^{\circ}/\text{s}$).
Cuturi et al. \cite{cuturi2014optic} provided evidence that visual sensitivity to radial optic flow is linked to vestibular sensitivity to forward and backward motion.
Furthermore, Clifford et al. \cite{clifford1999perception} found that the flow pattern has an effect on its perceived speed (with radial flow appearing faster than spiral and rotational flow patterns), and that the flow pattern type (complex motions consisting of local and global motion information versus simple motions that lacked any global motion information) did not have an effect on speed discrimination thresholds.

Notably, most prior work that measured radial optic flow thresholds used unnatural stimuli (e.g., random dot kinematograms), lacked stereo depth cues, or used a display with an extremely small field of view.
To the best of our knowledge, our work is the first to study sensitivity to radial optic flow using a rich, natural scene in a stereoscopic, wide field-of-view display, which is more representative of the perceptual experience when using a VR HMD.

\subsection{Blink-based Visual Suppression}
\label{subsec:blink_suppression}
Humans typically blink for anywhere in the range of 10 – 20 times per minute \cite{doughty2002further,leigh2015neurology} with a duration of $100-400$~ms \cite{holmqvist2017eye,adler1970physiology}.
During blinks, the observer is not able to see their surroundings since the eyelids are closed and prevent light rays from entering the eye.
However, visual suppression during blinks actually begins before a blink even begins and persists for a short duration after the blink ends \cite{volkmann1986human}.
Furthermore, blinks themselves usually go unnoticed due to decreased activity in the visual cortex (triggered by blinks) \cite{bristow2005blinking}.
In the context of virtual reality, blink suppression has been used to mask changes in the virtual environment to aid users in virtual locomotion \cite{langbehn2018blink,nguyen2018discrete} and user-object interaction \cite{zenner2021blink}.
\textcolor{black}{In the current work, we study to what extent blinks can be used to reduce an observer's sensitivity to radial optic flow artifacts, with applications to distortion correction for varifocal HMDs.}

\section{Methods}
\label{sec:methods}
To measure sensitivity to radial optic flow, we conducted a psychophysical experiment in which participants viewed a 3D virtual scene and were tasked with identifying when the size of the virtual image changed.
This section provides details on our experiment design, the equipment used, and the demographics of our participants.

\subsection{Equipment}
In order to avoid confounding factors related to headset fit and view-dependent effects from unconstrained head movements, and to ensure high accuracy of our eye tracking, we used the VR display simulator introduced by Guan et al. \cite{guan2022perceptual}.
The display simulator uses a 97'' high-speed OLED television (LG OLED97G2PUA) paired with shutter glasses to render a 3D image with a $125^\circ \times 94^\circ$ FOV.
The content was rendered at 60Hz per eye.
The television is mounted to a stand that allows it to be translated forward or backward within a 2m range.
The frames of the glasses are 3D printed, with lenses from Optoma ZD302 shutter glasses and a Tobii eye tracker installed on the interior of the frame.
We use a custom frame so that the glasses can be mounted to the display simulator's adjustable chin rest.
Attached on each side of the chin rest is a handlebar with two buttons to allow the user to input their response after a trial or to advance to the next trial.
The user sat in a height-adjustable stool and the vertical position and eye relief of the glasses could also be adjusted to ensure that the participant was comfortable during the experiment.
Note that this display simulator also supports rotational head movements, the use of a bite bar, and can render different types of simulated lens distortions, but we did not use any of these features for our experiment.
An image of the device can be seen in \autoref{fig:teaser}.

\subsection{Experiment Design \& Stimulus}
\textcolor{black}{
In this section, we provide details on the experiment design, including our psychophysical sampling paradigm, experiment flow, and properties of the image magnification stimulus.
Note that the terms ``image magnification,'' ``image size change,'' ``distortion,'' and ``radial optic flow'' all refer to the artifact produced by varifocal HMDs when the lens changes power, and we use these terms interchangeably throughout the paper.
}

\subsubsection{\textcolor{black}{Experimental Protocol}}
\label{subsubsec:exp_protocol}
\begin{figure*}[ht!]
    \centering
    \includegraphics[width=.95\textwidth]{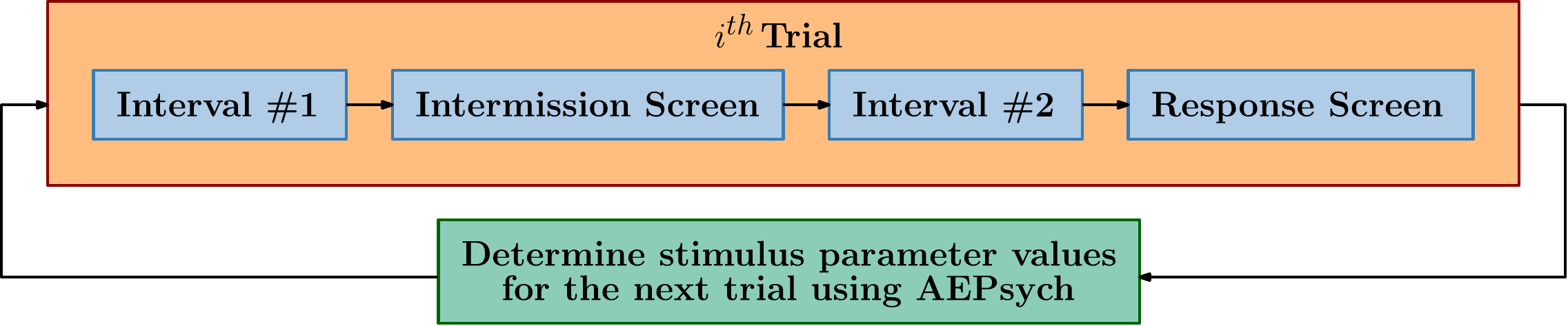}
    \caption{
    \textcolor{black}{A flowchart showing the procedure for a single trial in our 2IFC experiment. During each ``Interval'' block, the virtual scene was rendered on the display either with or without the radial optic flow distortion, which was triggered after the user initiated a blink (see \autoref{fig:teaser}). Only one of the two intervals had the distortion effect, chosen randomly for each trial. The intermission screen was used to mitigate any afterimage effects. On the response screen, participants pressed a button to indicate which of the two intervals they believed contained the radial optic flow effect. Before the next trial began, we used AEPsych \cite{owen2021adaptive} to assign the new values for the optic flow magnitude and delay between the blink and the optic flow onset in order to converge towards the participant's perceptual threshold for radial optic flow.}
    }
    \label{fig:exp_design}
\end{figure*}

\begin{figure}[t]
    \centering
    \includegraphics[width=.45\textwidth]{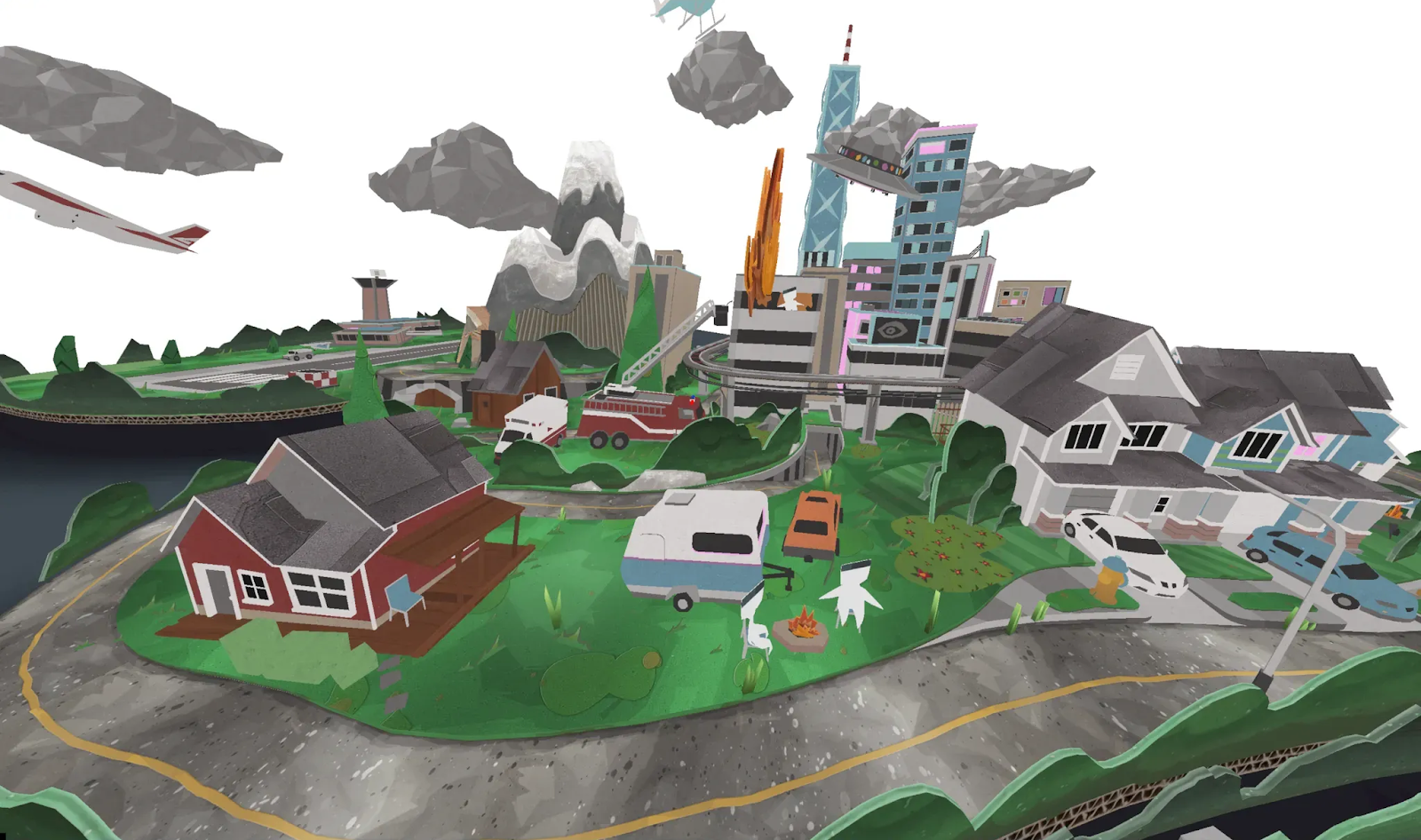}
    \caption{An image of Papertown, the 3D virtual scene that participants viewed. This scene was chosen because its rich and natural properties (many 3D objects with varying shape, texture, size, and color properties) make it representative of typical content that is viewed in VR HMDs. Note that participants viewed a still image of the scene, so it did not contain any dynamic elements.}
    \label{fig:papertown}
\end{figure}

To measure sensitivity to image size change, we conducted a 2-interval forced choice task (2IFC).
That is, for a single trial the participant was exposed to two intervals of the virtual content (a still image of the \textit{Papertown} scene shown in \autoref{fig:papertown}).
One interval had no change in image size, while in the other interval the stimulus \textcolor{black}{changed size. 
}
Each interval started with a presentation of the Papertown scene, followed by the appearance of a blink bue (a red circle) after $500$~ms.
Participants were instructed to blink at this moment.
Once the eye tracker detected the blink, the red circle disappeared and \textcolor{black}{stimulus was rendered depending on the interval (i.e., the virtual image size remained constant or was increased).}
From trial to trial, the order of the intervals was random and participants received feedback on the correctness of their response after each trial.
\textcolor{black}{An overview of a trial from our experiment is shown in \autoref{fig:exp_design}.}

The participant's task was to identify which of the two intervals contained the image size change.
When the magnitude of the \textcolor{black}{image magnification} was below the participant's detection threshold (i.e., the optic flow artifact was imperceptible), the participant had a $50\%$ chance of selecting the correct interval.
Conversely, as the magnitude of the size change increased, participants \textcolor{black}{were more likely to perceive} the stimulus and select the correct interval that contained the image si ze change.
We considered the perceptual threshold to be the point at which the participant had a $75\%$ chance of correctly identifying the stimulus interval.

Before beginning the experiment, participants were briefed on the the experiment task and the functionality of the different buttons to advance through the trials.
The experimenter also helped the participant adjust the height of the stool and the positioning of the chin rest and shutter glasses so that they were comfortable and were able to clearly see the display.
At the start of the experiment, participants completed 15 practice trials to get accustomed to the experiment flow and the controls.
Participants were free to take a break or end their participation at any time.
Participants were compensated with \$75 USD per hour upon completion of the experiment, and participation took about 1.5 hours.
\subsubsection{\textcolor{black}{Experiment Stimulus}}
\label{subsubsec:exp_stimulus}
\textcolor{black}{To understand the different parameters of image magnification that may affect perceptibility of the varifocal HMD's image distortion artifact, from trial to trial we adjusted the \textit{magnitude} of the image magnification and the \textit{delay} between the participant's blink onset and image magnification onset.
We chose to study these parameters in particular because they have important implications on the design and implementation of varifocal HMDs.
The magnitude of the magnification tells us how much room for error we have in the distortion correction pipeline, i.e. how good we need to be at eliminating the image magnification artifact (\autoref{subsec:VAC}).
The delay parameter tells us how fast the eye tracker and distortion correction processes need to be so that we can reliably mitigate the perceptibility the distortion artifact when the user blinks (\autoref{subsec:blink_suppression}).
}
The magnitude of the image magnification \textcolor{black}{ranged} between $0.5\%-3\%$ size change that happened over the course of  8.3 ms (one screen frame).
Additionally, the size change occurred on a random delay ($15-300\text{ ms}$) after blink onset.
\textcolor{black}{
Since blink suppression effects are known to change depending on \textit{when} the stimulus appears during the observer's blink, we varied the timing stimulus onset relative to the participant's blink onset to characterize how sensitivity to image magnification changes according temporally during blinks.
}

We used the AEPsych adaptive sampling toolbox \cite{owen2021adaptive} to modify the stimulus parameter values (magnification magnitude and delay after blink onset) since it has shown to be an efficient way of estimating perceptual thresholds in psychophysical studies \textcolor{black}{in which the perceptual effects of} multiple different stimulus parameters \textcolor{black}{are being studied} \cite{guan2022perceptual}.
Each experiment was run with 50 initialization trials \textcolor{black}{that uniformly sampled a subset of the parameter space that was likely to contain a participant's perceptual threshold ($0\%-2\%$ magnification magnitude and $0 - 150$ ms magnification delay).
The subset that we uniformly sampled was identified through extensive pilot testing, where we uniformly sampled the whole parameter space to learn which regions produced trials where the stimulus was consistently perceived (or not perceived); those trials correspond to stimuli that were consistently supra- or sub-threshold, and thus do not contribute much to AEPsych's ability to converge toward the perceptual threshold \cite{owen2021adaptive}.
This uniform sampling procedure acts like a prior that helps AEPsych more efficiently sample the parameter space to converge toward the participant's true threshold more reliably.
After the uniform sampling, we switched to AEPsych's adaptive sampling paradigm to collect data up to 250 trials.
}
\textcolor{black}{Note that since AEPsych uses an adaptive sampling paradigm to determine the parameter values, we did not divide the experiment into separate blocks with different (but constant within the block) parameter values as is often done in method of constant stimuli psychophysical experiments.
That is, on any given trial, the image magnification magnitude and delay parameters could take on any continuous value within their respective ranges.
This allows us to sample stimulus parameters very close to the participant's perceptual threshold, which is difficult to do with non-adaptive sampling paradigms since we do not know the participant's perceptual threshold before experimentation.
}

To implement the radial optic flow effect, we uniformly increased the width and height of the rendered image.
Note that by scaling the image size, the optic flow is not 100\% faithful to the kind of distortion artifact that is present in real varifocal HMDs, but at the small magnitudes we tested the difference between our \textcolor{black}{simulated magnification artifact} and the actual magnification artifact is negligible.
Furthermore, by scaling the entire image instead of changing the focal distance of the virtual camera used for rendering, we eliminated unwanted artifacts that participants could use to determine whether or not the optic flow was present.
In particular, scaling the entire image avoided \textcolor{black}{motion artifacts caused by} aliasing and the occlusion or disocclusion of some features in the scene (since it was a 3D scene).

\subsection{Participants}
A total of ten people (mean age 26.3 years, five female) successfully completed our user study.
Each participant was screened to ensure that the eye tracker could detect their blinks, they had stereo vision, normal or corrected-to-normal visual acuity, and were not color blind.

\section{Experiment Results}
\label{sec:results}

\subsection{Blink Duration}
\label{subsec:results_blinks}
\begin{figure}[t]
    \centering
    \includegraphics[width=.45\textwidth]{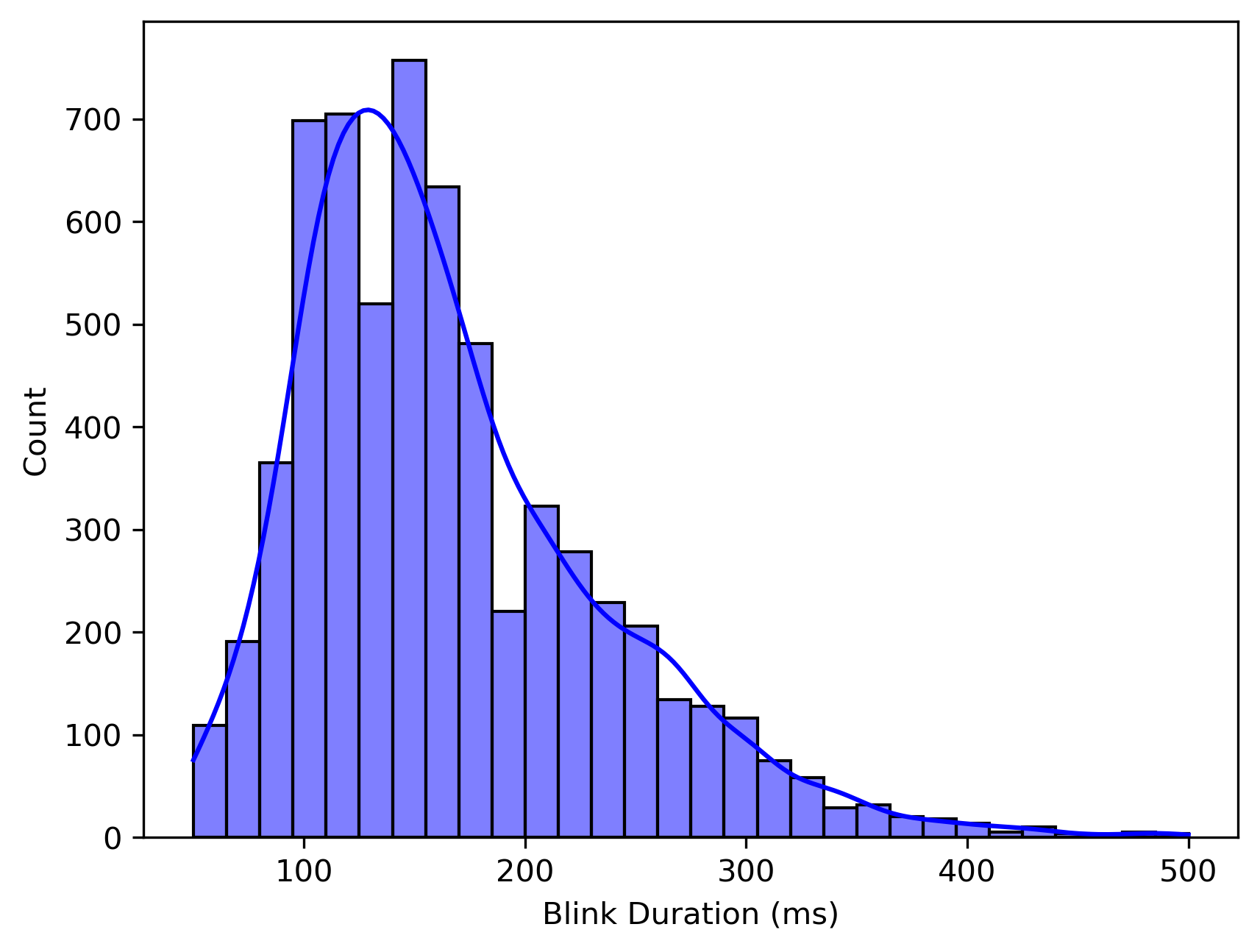}
    \caption{Distribution of blink durations (voluntary and involuntary) across all participants. The high concentration of blinks in the $100-200$ ms range aligns with prior research on blink durations~\cite{holmqvist2017eye}, which suggests that participants did not have abnormal blink behaviors in our experiment.}
    \label{fig:blink_distribution}
\end{figure}

\begin{figure}[t]
    \centering
    \includegraphics[width=.45\textwidth]{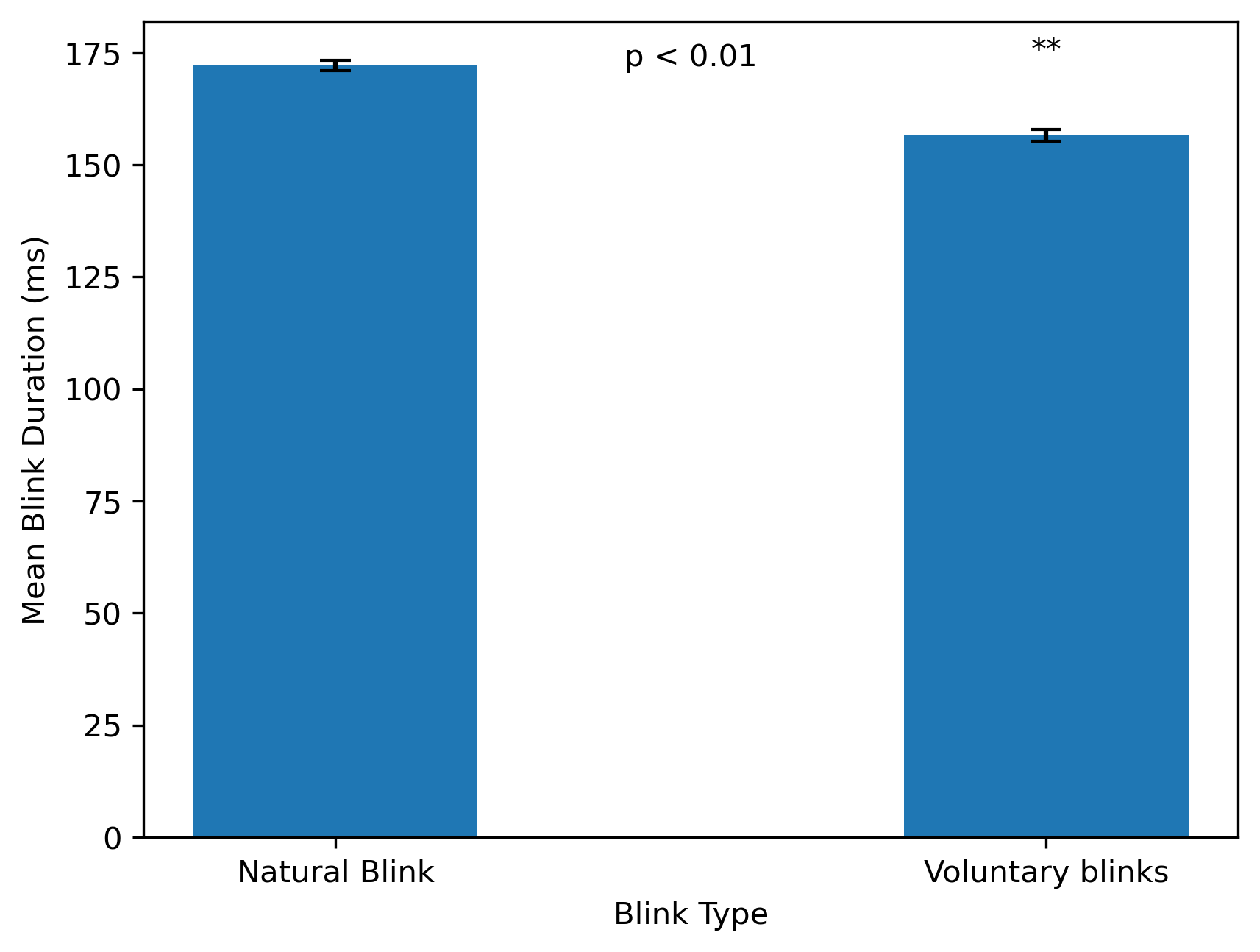}
    \caption{Average duration of voluntary and involuntary blinks. Involuntary (natural) blinks were significantly longer than involuntary blinks in our experiment. Although the absolute difference in duration between voluntary and involuntary blinks is not large, this statistical difference suggests that the engineering requirements for radial flow distortion may change depending on the type of blink a user initiates.}
    \label{fig:mean_blink_duration}
\end{figure}

To verify the validity of our blink detection algorithm, we recorded the duration of every blink (voluntary and involuntary) our participants initiated.
The distribution of blink duration is shown in \autoref{fig:blink_distribution}.
\textcolor{black}{We considered a blink to be voluntary if it was initiated when the red dot visual cue appeared at the start of an interval (which signaled to the participant that they should blink, see \autoref{subsubsec:exp_protocol}).
A blink was classified as involuntary if it occurred at any time other than shortly after the blink cue appeared.}
The median blink duration was $\sim \!\! 120$ ms, which is in line with prior research on blink duration \cite{holmqvist2017eye}.
Interestingly, we found that there was a significant difference in the duration of voluntary and involuntary blinks ($t = 8.93, p < .01$), with voluntary blinks being shorter than involuntary.
A bar chart comparing the duration of voluntary and involuntary blinks is shown in \autoref{fig:mean_blink_duration}.

\subsection{Radial Optic Flow Thresholds}
\label{subsec:results_thresholds}
\begin{figure*}[!ht]
    \centering
    \includegraphics[width=.985\textwidth]{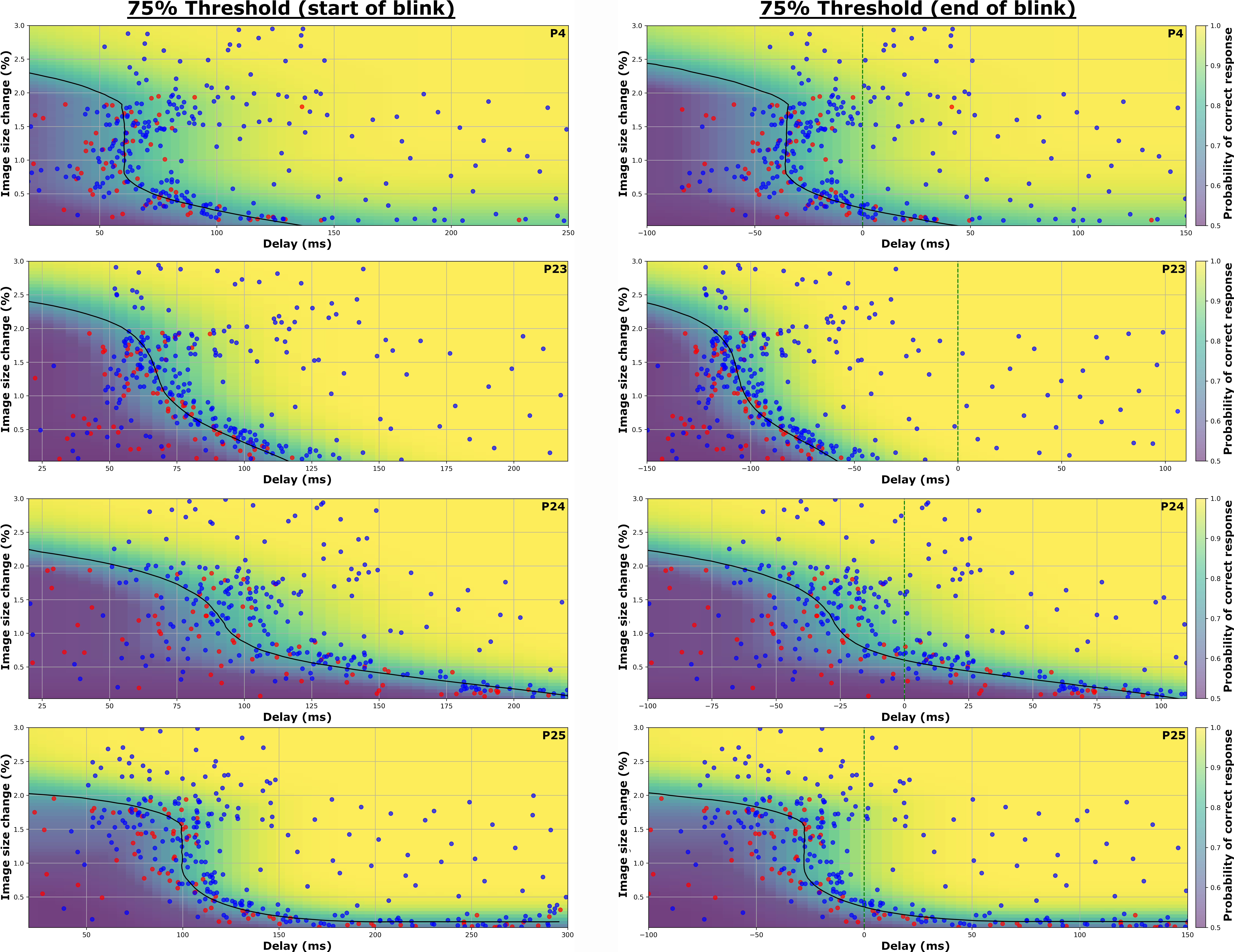}
    \caption{\textcolor{black}{Trial response data and} $75\%$ detection thresholds for four participants who completed our experiment. The solid black curve denotes the 75\% threshold, which is the combination of stimulus delay and magnitude that will yield a $75\%$ chance that the participant will be able to detect the change in image size. The blue and red dots are individual 2IFC trial responses, where blue is a correct response and red is an incorrect response. Note that the majority of trials are sampled near the threshold for each participant, which indicates that the AEPsych adaptive sampling was effective.
    \textcolor{black}{We only show four participants' data here to save space---the data are qualitatively very similar across all participants.}
    \textbf{Left~Column:} $75\%$ detection thresholds for participants when the stimulus was triggered at the \textit{start} of the voluntary blink (with varying delay). Visual sensitivity to radial optic flow is suppressed until $\sim \! 70$ ms after blink onset, after which point sensitivity begins to increase. \textbf{Right~Column:} $75\%$ detection thresholds for one participant when the stimulus was triggered at the \textit{end} of the voluntary blink (with varying delay). Visual sensitivity was suppressed towards the start of the blink ($\sim \! $ -100 ms) and, as the blink came to an end, the participant was easily able to detect even small image size changes of $0.5\%$.}
    \label{fig:individ_thresholds}
\end{figure*}

To find the $75\%$ detection threshold, we inverted the \textcolor{black}{fitted} AEPsych model~\cite{owen2021adaptive} and computed the radial optic flow magnitude and delay that produced a $75\%$ likelihood of detection (as predicted by the fitted model).
For model fitting, we enforced monotonicity with respect to the image size magnitude and delay in magnification.
For the analysis, we considered two ``blink events'' to study the temporal component of blink suppression to radial optic flow.
In particular, we considered image size changes that happened at the \textit{start} of a blink and at the \textit{end} of a blink.
The $75\%$ thresholds for four participants are shown in \autoref{fig:individ_thresholds}.

When considering radial flow that occurs at the start of a blink (\autoref{fig:individ_thresholds}, left column), image size changes of about $2.1\%$ and above were reliably detectable by the participant.
Additionally, we found that sensitivity begins to rapidly increase around $70$ ms after the start of a blink, which is around the time when the blink is coming to an end and the participant will be able to see the display again, although the eye may not be fully open.
As the delay increases, more and more of the stimulus is visible to the user after the blink has ended, and they are able to detect very small image size changes (around $0.05\% - 0.23\%$).
If the image size change occurs at the end of a blink (\autoref{fig:individ_thresholds}, right column), participants are able to reliably detect size changes below $0.5\%$.
This result matches the result found for radial flow that was triggered on a long delay ($100+$ ms) after the start of a blink, which confirms that participants are very sensitive to radial optic flow artifacts when they have a clear image of the virtual content.
Overall, the suppression effect for this participant began around $60$ ms after the start of a blink, and concluded roughly $70$ ms after the blink ended.

\begin{figure}[t]
    \centering
    \includegraphics[width=.45\textwidth]{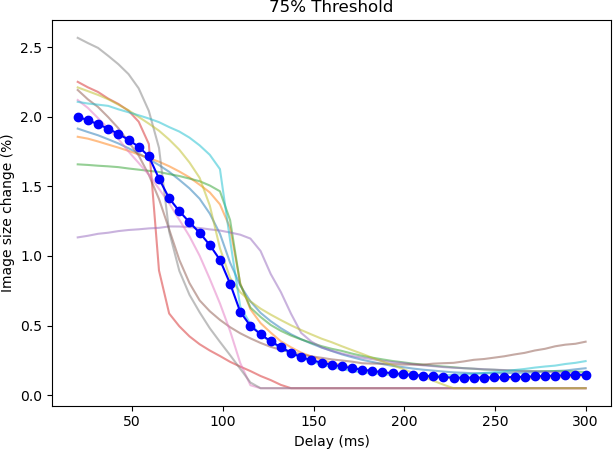}
    \caption{$75\%$ detection threshold for stimuli triggered at the start of a blink, averaged across ten participants (blue dotted curve). Overall, we see that visual sensitivity is suppressed until roughly $70$ ms after the blink begins, after which sensitivity rapidly increases and participants become extremely sensitive once the blink ends ($150$~ms and beyond).}
    \label{fig:threshold_start_avg}
\end{figure}

The average detection threshold across ten participants is shown in \autoref{fig:threshold_start_avg}.
The maximum image magnification we can hide in a blink without the user noticing is about $2\%$, with some inter-observer variability ($1.14\% - 2.54\%$).
The lower bound sensitivity after the blink suppression has resolved in the range of $0.05\% - 0.23\%$.
The difference between sensitivity at start of the blink and after the blink  is about one order of magnitude.

\begin{figure}[t]
    \centering
    \includegraphics[width=.45\textwidth]{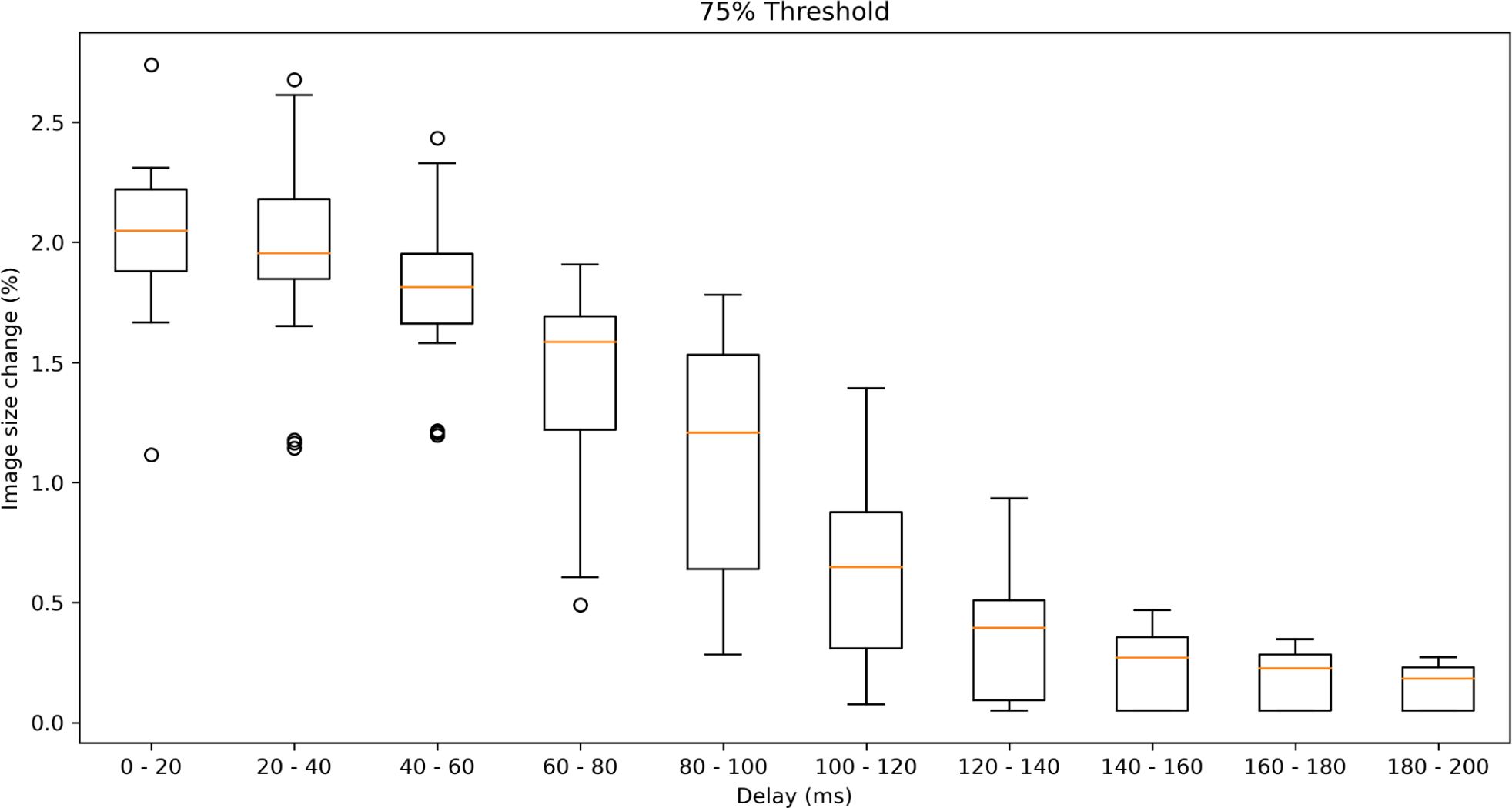}
    \caption{Sensitivity to radial optic flow triggered at the start of a blink, grouped by stimulus delay into buckets of $20$ ms. The general trend of the median sensitivity (orange horizontal lines) follows the average threshold shown in \autoref{fig:threshold_start_avg}. The long whiskers on the boxes in the $60-120$ ms range indicate that there is a large amount of inter-participant variability in sensitivity to radial optic flow toward the end of a blink.}
    \label{fig:buckets_boxplot}
\end{figure}

To get a better understanding of how sensitivity changes between participants over time, we grouped participant responses into $20$ ms buckets for the first $200$ ms after the start of a blink.
Results of this bucketing are shown in \autoref{fig:buckets_boxplot}.
Note that the plotted data come from trials where the stimulus was triggered at the start of the blink (e.g., the data shown in \autoref{fig:individ_thresholds} and \autoref{fig:threshold_start_avg}).
Results show a trend similar to that of \autoref{fig:threshold_start_avg}, but the large variation in the $60-120$ ms range indicates that there is a large amount of inter-participant variability in restoring visual sensitivity to image magnification.

\begin{figure}[t]
    \centering
    \includegraphics[width=.45\textwidth]{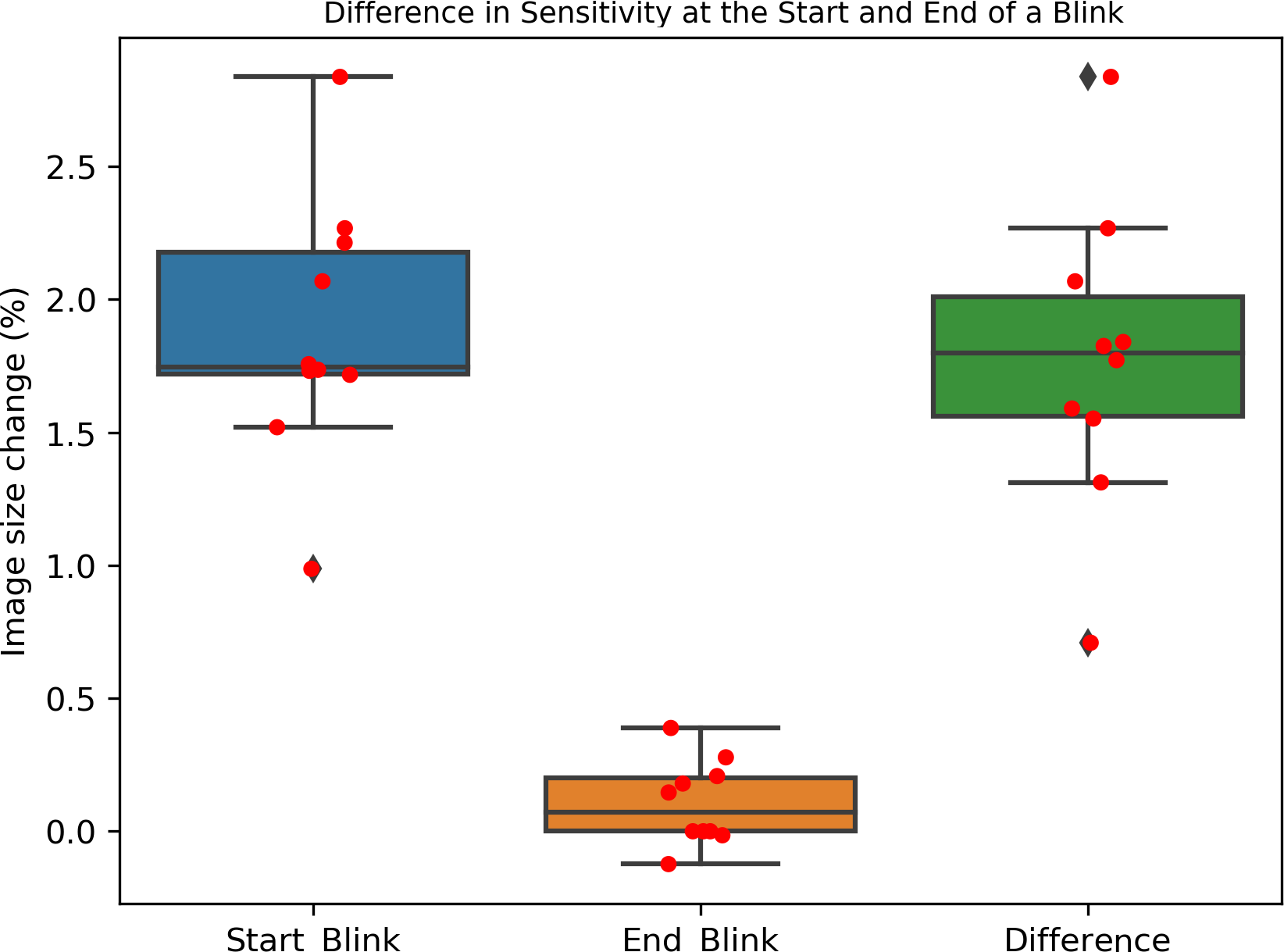}
    \caption{Difference between $0$ ms and $200$ ms sensitivity per user for stimuli triggered at the start of a blink event, end of a blink event, and the difference in thresholds between start and end of blink events. Red dots denote the raw threshold data for each participant. This plot highlights the impact of blink suppression (green box): When blink suppression is active (``Start Blink'' condition), participants are $10\times$ less sensitive to changes in image size compared to when suppression is inactive (``End Blink'' condition).}
    \label{fig:start_vs_end_difference}
\end{figure}

Finally, \autoref{fig:start_vs_end_difference} shows the distributions in each participant's $75\%$ threshold for stimuli triggered at the start of a blink, the end of a blink, and the difference in image size change between the start and end thresholds.
This plot highlights the stark change in sensitivity as a function of image size change relative to blink, i.e. the decrease in sensitivity afforded by blink suppression.

\section{Discussion}
\label{sec:discussion}

The results of our experiment showed that blinks can be used to reliably mask the radial optic flow artifact that is associated with changes in focal power of varifocal display systems.
In particular, we found that image size changes of $2.032 \pm 0.45\%$ can be hidden during blinks, and that visual sensitivity returns roughly $70$~ms after a blink begins (as detected by our eye tracker, which has its own inherent latency).
Once visual sensitivity has returned, we found that participants were very accurate in detecting changes in image size; participants were able to detect images that increased in size by just $0.05\% - 0.23\%$.

\paragraph{\textcolor{black}{\textbf{Varifocal HMD Design}}}
Our results have several important implications for the design and implementation of varifocal HMDs.
First, the high perceptual sensitivity when users' view of the virtual image changes in this manner clearly indicates that margin for error on imagine magnification correction is \textit{very} small.
That participants were able to detect even $0.05\%$ changes highlights that we are very sensitive to radial optic flow in near-eye displays and distortion correction needs to be very accurate to deliver a comfortable experience for users.
Currently, the two main approaches to varifocal HMDs are discrete and continuous varifocal.
With discrete varifocal systems, the focal plane changes almost instantly \cite{zhan2020multifocal}, which is perceived as a very short duration image magnification for the user. This is nearly identical to our experiment, where the image size change occurred over the course of one frame ($\sim \!\! 8$~ms). 
In such systems, the instantaneous image size change could theoretically be masked by a blink (\autoref{subsec:results_thresholds}), which significantly decreases the burden on the rendering system for distortion correction.
In a continuous varifocal design, the virtual image distance changes smoothly over the course of $200-700$~ms (e.g., \cite{zhao2023retinal}).
In these systems, the retinal image size changes more slowly over a longer duration.
Since our experiment did not directly measure thresholds for retinal flow velocity, it is unclear whether continuous varifocal systems can leverage blinks to reduce the impact of image magnification artifacts.
However, the slower changes in retinal image size will likely reduce the salience of the uncorrected distortion---both in terms of perceptibility and acceptability.

\paragraph{\textcolor{black}{\textbf{Distortion Correction}}}
Aside from implications on the viability of different varifocal HMD designs, our results also have implications for the engineering requirements for the software-based distortion correction.
Distortion correction in software is typically done by modeling the optics of the HMD and applying corrective distortions to the rendered image to yield an undistored image when viewed through the HMD's lens \cite{rolland1993method,watson1995using,bax2002real,robinett1992computational}.
However, the perceived distortion depends not only on the lens properties but also on the position of the eye relative to the lens \cite{jones2015correction}.
Thus, in order to sufficiently correct any distortion artifacts (like the radial optic flow artifact studied in this work), the system should ideally have an accurate measurement of the position of the user's eye within the HMD eye-box.
Given that results \textcolor{black}{presented in this work} show that the timing of the distortion correction relative to eye blinks is crucial for masking any artifacts, our results also have implications on the minimum requirements for eye tracking latency and distortion mesh refresh rates.
In particular, eye tracking and distortion mesh updates should be fast enough to accurately measure the eye's position, detect a blink, and update the distortion mesh all within $\sim \!\! 70-100$~ms in order to minimize the perception of radial optic flow artifacts.

\paragraph{\textcolor{black}{\textbf{Voluntary Versus Involuntary Blinks}}}
In our experiment, we also found that the duration of voluntary and involuntary blinks was different (\autoref{subsec:results_blinks}).
In particular, we found that involuntary blinks were about $20$~ms longer than voluntary blinks.
In terms of distortion correction, this is favorable since it suggests that involuntary blinks which occur naturally as the user engages with virtual content are more useful for masking the radial optic flow artifact.
Additionally, this suggests that varifocal systems may benefit from VR experiences that intentionally induce blinks in the user \cite{zenner2023induce}, since this can be used to increase the frequency of natural blinks which can be used for artifact masking.

\paragraph{\textcolor{black}{\textbf{Individual Differences}}}
\textcolor{black}{
Individual differences are a well-known phenomenon in perception \cite{witkin1949nature,lafer2015striking}.
In our experiment, we found some differences in thresholds between participants.
There may be implications of individual differences on the implementation of distortion correction for varifocal HMDs.
In particular, the timing of when to initiate a change in lens power relative to the detection of a blink may change from user to user.
For example, for the threshold curves shown in \autoref{fig:threshold_start_avg}, the start of the sensitivity increase varies within the $60-120$~ms range after blink onset.
Despite the differences in temporal properties of the blink suppression observed in our experiment, the baseline sensitivity after participants' blinks end is relatively constant across all participants.
This means that while individual differences may have implications for the temporal properties of the implementation distortion correction, the requirements for the quality of the distortion correction (i.e., the amount of distortion that must be corrected) is relatively constant (and high) across all participants.
These results may allow for more relaxed constraints on eye tracker and blink detection latency, but likely will not have any major implications for the implementation of the distortion correction algorithm itself, which must be very accurate in order to provide a pleasant user experience.
}

\paragraph{\textcolor{black}{\textbf{Blinks and Vergence Eye Movements}}}
\textcolor{black}{
It should be noted that, in an effort to keep the experiment simple and to measure baseline, ``worst-case'' thresholds, participants did \textit{not} make vergence movements during our experiment.
Varifocal HMDs specifically designed to change the focal power of the HMD's lens when the user changes their vergence depth in VR, so using blinks to mask the distortion artifact would yield the greatest benefits if blinks occurred simultaneously with vergence movements.
Indeed, Rambold et al. \cite{rambold2002effects} observed that most blinks in their experiment occurred at the start of or during vergence eye movements, and rarely occurred before.
However, it is important to note that blinks last about 120 ms (\autoref{fig:blink_distribution}) but vergence movements have a much longer duration of $\sim \!\! 330$ ms \cite{tyler2012analysis}.
Therefore, while the data from prior work \cite{rambold2002effects} suggest that blinks are sufficiently concomitant with vergence movements for us to take advantage of blink suppression for varifocal distortion masking, the engineering requirements for distortion correction are constrained by the duration of the user's blinks since they are typically much shorter in duration than vergence movements.
Nevertheless, additional studies should be conducted to better understand the relationship between blinks and vergence movements and the implications for blink suppression in \textit{head-mounted} varifocal displays during more representative use-cases of VR.
}

\paragraph{\textcolor{black}{\textbf{Post-Blink Sensitivity Suppression}}}
Finally, \textcolor{black}{one must consider} that unlike most blink suppression research, we studied sensitivity to an artifact that persists even \textit{after} a blink has concluded.
Thus, in this work we were able to characterize how visual sensitivity changes not only during stimulus exposure, but also after a blink.
Our results showed that suppressed sensitivity to radial optic flow still exists for a short duration after a blink ends (about $75$~ms).
This existence of post-blink suppression may indicate a similar effect for other artifacts, and we believe it would be worthwhile for future research to investigate this.



\section{Conclusion, Limitations, \& Future Work}
\label{sec:conclusion}

In this work, we measured perceptual sensitivity to radial optic flow distortion effects in near-eye stereoscopic displays.
Radial optic flow is of interest because it is an artifact that is inherent to varifocal display systems; when the focal power of the display changes, the virtual image increases or decreases in size, which the user perceives as a radial optic flow effect.
By measuring how sensitive users are to this flow artifact and to what extent this sensitivity can be decreased through blink suppression, we gain an understanding of how accurate the distortion-correction systems for varifocal HMDs must be.
Our results showed that blinks can be leveraged to mask radial optic flow effects.
Compared to flow effects that occurred $\sim \!\! 70$~ms after blink onset, we found that visual sensitivity was decreased tenfold at the start of a blink.
We also found that pre- and post-blink sensitivity was, on average, $\sim~\! 0.15\%$ image size change when no blinks were present and increased to $\sim~\!\! 1.5-2.0\%$ during blinks. 

There are some limitations to our work.
First, we only tested sensitivity to static virtual content.
It is possible that dynamic virtual content, as is common in virtual experiences, may serve as an additional ``distraction'' that could further reduce the user's sensitivity to radial optic flow.
Additionally, the thresholds reported in our study are likely conservative estimates since participants were specifically instructed to try to perceive the artifact and they sat still for the duration of the experiment.
In normal VR settings, the user is likely to be moving around and interacting with their virtual surroundings, which may have additional effects on their ability to perceive radial flow effects.
Our display system had a fixed distance between the user and they display, which means that the user did not undergo any change in accommodation before, during, or after the image changed in size.
Since a change in accommodation causes momentary blur of the retinal image, it is possible that this could impact the user's sensitivity to radial optic flow.
\textcolor{black}{
Furthermore, it is possible that participants' blinking behavior during our experiment was different from normal blinking behavior due to the increased amount of attention that participants devote towards blinking \cite{ang2020boosted,hoppe2018humans,nakano2009synchronization}, which may have affected our results.
}

Future work should study observers' sensitivity to radial optic flow during vergence movements.
Since the image size change effect occurs when the user shifts their gaze to look at an object at a different virtual distance (i.e., they do a vergence movement), observers may have reduced sensitivity (compared to normal viewing with no eye movements or blinks) due to vergence suppression effects \cite{manning1984vergence,hung1989suppression,hung1990suppression}.
Additionally, sensitivity to radial optic flow as a function of the observer's level of cognitive load should be measured, since this is more representative of the situations in which users are likely to experience this flow artifact when using a varifocal HMD and prior work has shown that cognitive load can affect motion perception in VR \cite{nguyen2020effect,williams2019estimation}.

\acknowledgments{%
The authors thank Joel Hegland and Olivier Mercier for hardware and software support for the stereoscopic display.
Niall L. Williams was supported in part by the Link Foundation Modeling, Simulation, \& Training Fellowship.
}

\bibliographystyle{abbrv-doi-hyperref}

\bibliography{template}

\end{document}